# Cr (VI) biosorption at different Na (I) concentrations after Zn(II) uptake during *Arthrobacter species* growth

E.Gelagutashvili, O.Rcheulishvili, A.Rcheulishvili , M.Janjalia

*Iv. Javakhishvili Tbilisi State University*
*E. Andronikashvili Institute of Physics*
*0177, 6, Tamarashvili St.,*
*Tbilisi, Georgia*

## Abstract

The biosorption of Cr(VI) *Arthrobacter species (Arthrobacter globiformis 151B and Arthrobacter oxides 61B*) after growth in the presence of Zn (II) ions at different Na (I) concentrations was studied application using dialysis and atomic absorption analysis.

It was shown that after absorption of Zn(II) during growth of types of *Arthrobacter* the biosorption of Cr (VI) with types *Arthrobacter (Arthrobacter globiformis 151B, Arthrobacter oxidas 61B*) increases at various concentration of Na (I) comparative without Zn(II), at identical concentration of Na (I). It is seen, that zinc(II) ions promote for Cr(VI) part of the active centers of *Arthrobacter species* and Na(I) ions already has relatively little effect on them. However, this change is more pronounced for Cr (VI) _*Arthrobacter globiformis 151B* + Zn (II) than for Cr (VI) _*Arthrobacter oxides of 61B* + Zn (II).

## Introduction

Hexavalent chromium poses a risk due to its carcinogenic and high toxicity properties to living organisms. Cr(VI) to induce acute and chronic toxicity, neurotoxicity, genotoxicity, carcinogenicity, and immunotoxicity in humans and animal [1,2]. Heavy metals are among the most common pollutants found in the environment. Health problems due to the heavy metal pollution become a major concern throughout the world, and therefore, various treatment technologies are adopted to reduce or eliminate their concentration in the environment. Biosorption can be used for detoxification of heavy metals in industrial effluents [3]. In [4] were discussed different types of kinetic, equilibrium, and thermodynamic models used for HM treatments using different bacterial species, as well as biosorption mechanisms along with desorption of metal ions and regeneration of bacterial biosorbents. Based on the results, it seems that most bacteria have the potential for industrial applications for detoxification of HMs [4]. Biosorption has emerged as an alternative to conventional effluent treatment methods. Compared with conventional methods biosorption technology possesses several advantages: low operating cost, high efficiency in detoxifying heavy metals that have lower concentrations, less amount of spent biosorbent for final disposal, and no nutrient requirements[5]. Biosorption can be used for removal of Cr (VI) from industrial effluent, in order to avoid possible acute and chronic chromium poisoning of living beings. The result indicated that the chemical interactions such as ion-exchange between the hydrogen atoms of carboxyl (-COOH), hydroxyl (-OH) and amine (-$NH_2$) group of biomass and metal ions are mainly involved in biosorption of Cr(VI) onto *A. niger* surface. The changes in the functional groups and the surface properties of pretreated fungal biosorbent were confirmed by FTIR spectra [6]. Gram-positive *Arthrobacter species* bacteria can reduce Cr(VI) to Cr(III) under aerobic growth and there is a large interest in Cr- reducing bacteria. A culture of *Arthrobacter sp.* was tested for its ability to sorb copper, cadmium, and iron ions and chemical modeling of experimental data was used to interpret the mechanism of biosorption [7]. The biosorption of metals depends on external factors, such as other metal ions in the solution, which may appear to be competitors. In work [8] has shown that the chromoreducting activity of *Arthrobacter rhombi-RE* is dependent on other metal ions. Cd (II), Hg (II) ions inhibited the enzyme activity. The exact mechanism by which microorganisms take up the metal is relatively unclear. In our previous works [9, 10] were investigated biosorption of Cr(VI) and Cr(III) with *Arthrobacter species* and influence Na(I) ions on that.

In this paper were studied the biosorption of Cr(VI) by *Arthrobacter* species (*Arthrobacter globiformis 151B and Arthrobacter oxides 61B*) after during growth in the presence of Zn(II) ions at room temperature at various Na(I) concentrations simultaneous application dialysis and atomic absorption analysis.

## Materials and Methods

The other reagents were used: NaCl, $K_2CrO_4$, $ZnCl_2$ (Analytical grade). *Arthrobacter* bacterials were cultivated in the nutrient medium loaded with concentration of Zn (50mg/l). *Arthrobacter* species cells were centrifuged at 12000 rpm for 10 min and washed three times with phosphate buffer (pH 7.0). The centrifuged cells were dried without the supernatant solution until constant weight. After solidification ( dehydrated) of cells (dry weight) solutions for dialysis were prepared by dissolving in phosphate buffer. This buffer was used in all experiments. A known quantity of dried bacterium suspension was contacted with solution containing a known concentration of metal ion. For biosorption isotherm studies, the dry cell weight was kept constant (1 mg/ml), while the initial chromium concentration in each sample was varied in the interval ($10^{-3}$ -$10^{-6}$ M). All experiments were carried out at ambient temperature. Metal was separated from the biomass with the membrane, which thickeness was 30μm Visking (serva) and analyzed by an atomic absorption spectrophotometer „Analyst-900'' (Perkin Elmer) $\lambda_{Cr}$=357.9 nm wavelength. Dialysis carried out during 72 h. Concentration of Na(I) was 2mM, 20 mM , 50mM. The isotherm data were characterized by the Freundlich [11] equation, which by us in analogue cases were discussed in work [9].

## Results and Discussions

The biosorption of Cr(VI) by *Arthrobacter* species ((*Arthrobacter globiformis 151B and Arthrobacter oxidas 61B* ) after during growth in the presence of Zn(II) ions at room temperature at various Na(I) concentrations simultaneous application dialysis and atomic absorption analysis were studied. Freundlich parameters evaluated from the isotherms with the correlation coefficients are given in table 1 of Cr(VI) ions for two kinds of *Arthrobacter* (*Arthrobacter globiformis 151B and Arthrobacter oxidas*) after during growth in the presence of Zn(II) ions at various Na(I) concentrations. The correlation coefficient value for different sets of experiments were found to be higher than 0.92. For comparison in table 1 are shown also biosorption constants at various Na(I) concentrations and without Na(I) from our recent works[9.10]. As seen from table 1, the change in sodium concentration strongly affects Cr(VI)-*Arthrobacter* species complexes as different concentrations Na(I) without Zn(II) ions, as at various concentrations Na(I), when Zn(II) ions are loaded in medium at *Arthrobacter* species cultivation and significantly difference from biosorption *Arthrobacter* species without uptake Zn(II) ions during cultivation. In particular, as the concentration of sodium increases, the binding constant decreases in all cases. Comparative biosorption characteristics for Cr(VI) *Arthrobacter* species+Zn(II) shows, that with increasing Na(I) concentrations , biosorption decreases in both cases. (Decreases from 5.9 x$10^{-4}$ to 4.9 x$10^{-4}$ in the case of Cr (VI) _*Arthrobacter oxidas 61B,* and for Cr (VI) -*Arthrobacter globiformis* 151B 6.8 x$10^{-4}$ to 4.6 x$10^{-4}$), but after absorption of Zn(II) during growth of types of *Arthrobacter* species the biosorption of Cr (VI) with types *Arthrobacter (Arthrobacter globiformis 151B, Arthrobacter oxidas 61B*) is more at various concentration of Na (I) comparative without Zn(II), at identical concentration of Na (I). Biosorption parameters for Cr(VI) *Arthrobacter* species shown, that in the presence Zn(II) ions more decrease in bioavailability has been observed experimentally for Cr(VI)- *Arthrobacter globiformis* as compared with Cr(VI)- *Arthrobacter oxidas*. It is seen also from table 1, that at any concentrations of Na(I) ions and without it, the biosorption constant as *Arthrobacter oxidas,* as *Arthrobacter globiformis* is greater in the case, when Zn(II) ions loaded during cultivation, than without it. After comparing the results, it is seen that for *Arthrobacter oxidas* the biosorption constant without Na(I) is 1.8 times greater than at 50 mmol Na(I). The same results only with cultivation in the presence of Zn(II) ions is 1.35. The relation of results for *Arthrobacter oxidas* at 2 mM Na(I) and 50mM Na(I) makes 1.5, when after Zn(II) uptake during *Arthrobacter oxidas* growth the same relation is 1.2. The same relations for *Arthrobacter globiformis* makes 3.6 and 1.8, and 2.2 and 1.5 respectively. These means that zinc(II) ions promoted for Cr(VI) part of the active centers of *Arthrobacter species* and Na(I) ions already has relatively little effect on them. However, that change is more pronounced for Cr(VI) _*Arthrobacter globiformis 151B+Zn(II)* compared to Cr (VI) _*Arthrobacter oxidas 61B+ Zn(II)*.

Table 1. Biosorption parameters for Cr(VI)_*Arthrobacter* species during growth in the presence of Zn(II) at various Na(I) concentration

| | [Na$^+$], mM | Biosorption constant, Kx10$^{-4}$ | Absorption capacity, n | Correlation coefficient R$^2$ |
|---|---|---|---|---|
| Cr(VI) *Arthrobacter oxidas 61B* [9,10] | 50 | 2.51 | 1.92 | 0.98 |
| | 20 | 3.23 | 1.58 | 0.99 |
| | 2 | 3.8 | 1.03 | 0.93 |
| | without Na$^+$ | 4.6 | 1.25 | 0.98 |
| Cr(VI) *Arthrobacter oxidas* +Zn(II) | 50 | 4.9 | 1.67 | 0.93 |
| | 20 | 5.2 | 0.9 | 0.96 |
| | 2 | 5.9 | 1.45 | 0.95 |
| | without Na$^+$ [9] | 6.6 | 1.08 | 0.98 |
| Cr(VI)- *Arthrobacter globiformis 151B* [9.10] | 50 | 0.95 | 1.05 | 0.97 |
| | 20 | 1.59 | 1.2 | 0.95 |
| | 2 | 2.09 | 1.49 | 0.99 |
| | without Na$^+$ | 3.4 | 1.35 | 0.96 |
| Cr(VI)- *Arthrobacter globiformis 151B+Zn(II)* | 50 | 4.6 | 1.12 | 0.97 |
| | 20 | 5.0 | 0.59 | 0.97 |
| | 2 | 6.8 | 0.85 | 0.98 |
| | without Na$^+$ [9] | 8.1 | 1.19 | 0.96 |

Thus, It was shown that after absorption of Zn during growth of types of *Arthrobacter* the biosorption of Cr (VI) with types *Arthrobacter* (*Arthrobacter globiformis 151B, Arthrobacter oxidas 61B*) increases at various concentration of Na (I) comparative without Zn, at identical concentration of Na (I). As for Absorption capacity, *n* in all discussed cases it changes from 0.59 to 1.92. Generally speaking, the maximum biosorption capacities for all the studied heavy metals and types of brown algae are quite high, ranging from 0.39 to 1.66 mmol/g. Most of sorbents can have the $q_{max}$ above 0.8 mmol/g [12].

Due to the repulsive electrostatic interactions, Cr(VI) anion species are generally poorly adsorbed by the negatively charged soil particles and can move freely in the aqueous environments[13]. By modeling the experimental results obtained, the energy parameters defined for the two types of bacteria can be used to explain the mechanisms of biosorption.